\documentclass[authoryear,preprint,review,12pt]{elsarticle}

\usepackage{amsmath}
\usepackage{amssymb}
\usepackage[nomarkers,figuresonly]{endfloat}

\newcommand{\Dts}[1]{ \frac{D^\sharp {#1}}{D t} }
\newcommand{\ddt}[1]{ \frac{\partial {#1}}{\partial t} }
\newcommand{\ddx}[1]{ \frac{\partial {#1}}{\partial x} }
\newcommand{\ddy}[1]{ \frac{\partial {#1}}{\partial y} }
\newcommand{\ddz}[1]{ \frac{\partial {#1}}{\partial z} }
\newcommand{\ddtb}[1]{ \frac{\partial {#1}}{\partial \tilde{t}} }
\newcommand{\ddxb}[1]{ \frac{\partial {#1}}{\partial \tilde{x}} }
\newcommand{\ddyb}[1]{ \frac{\partial {#1}}{\partial \tilde{y}} }
\newcommand{\ddb}[1]{ \frac{\partial {#1}}{\partial \tilde{b}} }
\newcommand{\ol}{ \overline }

\newcommand{\Ee}{ \mathcal{E}' }
\newcommand{\IEe}{ \langle\Ee\rangle_z }
\newcommand{\Ke}{ \mathcal{K}' }
\newcommand{\Pe}{ \mathcal{P}' }
\newcommand{\Km}{ \ol{\mathcal{K}} }
\newcommand{\Pm}{ \ol{\mathcal{P}} }

\makeatletter
\def\ps@pprintTitle{%
 \let\@oddhead\@empty
 \let\@evenhead\@empty
 \def\@oddfoot{September-2017}%
 \let\@evenfoot\@oddfoot}
\makeatother

\begin{document}

\begin{frontmatter}



\title{A Prognostic, One-Equation Model of Meso-Scale Eddy Momentum Fluxes}


\author{Juan A. Saenz\corref{cor1}}

\author{Todd D. Ringler\corref{cor2}}

\address{Los Alamos National Laboratory}

\cortext[cor1]{juan.saenz@lanl.gov}
\cortext[cor2]{ringler@lanl.gov}
\cortext[cor3]{LA-UR-20-30345}

\begin{abstract}
We present a prognostic, one-equation model for eddy-mean flow force interactions to parameterize the divergence of the Eliassen-Palm flux tensor (EPFT) that arises from thickness-weighted averaging (TWA) the hydrostatic Boussinesq equations. The TWA system of equations does not invoke any approximations beyond the ones for which the hydrostatic Boussinesq equations are valid, and therefor constitutes a mathematically consistent framework with clear physical interpretations that we use to develop our model. This model is intended for the adiabatic interior of zonally symmetric flows, in the absence of topographic features that break zonal symmetry, where the two terms corresponding to eddy interfacial form drag in the EPFT dominate forces that result from its divergence. We model eddy interfacial form drag terms for vertical flux of horizontal momentum in the EPFT using the gradient hypothesis, which approximates these fluxes as the product of an eddy viscosity and the vertical gradient of horizontal momentum. We use mixing length theory to relate the viscosity to an eddy length scale and an eddy velocity, which in turn is proportional to the eddy energy, or the sum of eddy kinetic and eddy potential eddy in the TWA system. To model the eddy length scale, we approximate it as the first Rossby radius of deformation, which we calculate as a function of the mean flow. In our approach, we use a prognostic equation for vertically integrated eddy energy at each horizontal location, which we derive from the TWA framework, and then simplify to the flows of interest by ignoring transport, redistribution and diabatic terms. The prognostic vertically integrated eddy energy is projected onto the water column using an approximation to the eigenvalue of the first baroclinic mode to obtain the eddy energy at each vertical position. Thus, the eddy viscosity in our model has horizontal as well as vertical structure. We diagnosed the model equations in a high resolution, eddy resolving numerical simulation of a zonally re-entrant channel representative of the Southern Ocean. We have implemented the model parameterization in an ocean model and tested it to simulate a parameterized simulation of this flow.
\end{abstract}


\end{frontmatter}

%\linenumbers


%%%%%%%%%%%%%%%%%%%%%%%%%%%%%%%%%%
\section{Introduction}
\label{s:introduction}

Interactions between eddies and the mean flow regulate the momentum balance of ocean currents, and stirring by meso-scale eddies affects mixing and transport of tracers in the ocean, such as heat, salt and CO$_2$ among others.
Meso-scale eddies contain most the ocean kinetic energy.
However, climate models do not routinely use grids in the ocean with resolutions small enough to resolve ocean meso-scale eddies, so eddy processes have to be modeled.

Since its inception, the \citet[][GM]{gent_mcwilliams_1990} parameterization has played a pivotal role in modeling ocean circulation and global climate \citep{dababasoglu_etal_1994}.
In the GM parameterization, the effects of eddies on the Eulerian mean flow Boussinesq equations is represented by the eddy flux of layer interface height, or eddy form drag, which is interpreted as a quasi-adiabatic eddy induced transport velocity that advects tracers in the Eulerian-averaged governing equations.
The GM parameterization is an algebraic model, in which the modeled eddy fluxes are determined by the Eulerian mean state variables through an algebraic function.
This function has a one parameter, the `GM kappa' $\kappa_{GM}$, which is often assumed to be constant. 
Given the algebraic nature of the GM parameterization, many other complex features of ocean turbulence are not representable by GM, such as history effects, or time correlations, among others.
This is not a problem in simple flows with zonal symmetry, where history effects are not important.
%Work has been done to model $\kappa_{GM}$ as a function of stratification, giving it a vertical structure.

An alternative approach is to use the thickness-weighted averaged (TWA) Boussinesq equations, and represent eddy-mean flow interactions in the momentum equations as the divergence of the Eliassen-Palm (momentum) flux tensor (EPFT), or eddy forces on the mean flow.
Similar approaches have been suggested before in a number of different studies \citep{visbeck_etal_97, ferreira_marshall_2006, eden_greatbatch_2008}, which differ from the TWA formulation by \cite{deszoeke_bennett_1993} in a number of ways \citep[see][]{young_2012, maddison_marshall_2013}.
In the TWA approach, the prognostic variables in the hydrostatic Boussinesq equations are weighted by layer thickness (per unit buoyancy) and averaged over an ensemble of realizations, without incurring in any approximations or assumptions beyond those in the hydrostatic Boussinesq equations.
The result is an exact set of equations with the same mathematical form as in the un-averaged system, with additional terms containing the effects of unclosed eddy correlations, which appear in conservative flux form.
For example, in the momentum equations, the only additional terms that appear are eddy forces on the mean flow, and have the form of the divergence of the EPFT, a momentum flux tensor.
The terms in the EPFT represent eddy shear, eddy potential energy and eddy form drag, the latter being the focus of most modeling in the past few decades, having connections to the bolus velocity in GM \citep[e.g.][]{gent_mcwilliams_1990, greatbatch_lamb_1990, saenz_etal_2015a}.

In the presence of topography, as in the Southern Ocean around the Kurguelan Plateau and Drake Passage, among other topographical features, shear and barotropic instabilities lead an intensification of eddy activity.
Thus, other processes, aside from eddy form drag, are important in these regions.
Algebraic models are inherently limited in their capacity to model eddy-mean flow interactions in these flows, as the history of the flow is ignored.
For this reason, the strong mathematical footing and clear physical interpretation make the TWA framework a good candidate to model the most important aspects of the complexity or eddy-mean flow interactions.
With the TWA, one can formulate prognostic equations for the EPFT, or related terms, and model them by making specific assumptions.
For example, one could envision having a system of prognostic closure equations comprised by a partial differential equation (PDE) governing the EPFT terms, along with other equations needed to close it.
This way, we could keep track of the history of EPFT term transport, and transfers among them.
To accomplish this, we need diagnostics of the EPFT in models where important processes are represented, e.g. the flow over topography, in order to identify the most important terms in the EPFT, and thus inform us on the complexity of the prognostic models.

Here, as a first step towards moving beyond algebraic parameterizations using the TWA framework, we develop a one-equation turbulence closure for eddy-mean flow interactions in zonally symmetric flows.
In this model, we only account for the vertical flux of horizontal momentum, which is modeled using the gradient diffusion hypothesis.
The resulting eddy viscosity is represented using mixing length theory.

We proceed as follows. In section \ref{s:govEqns} we present the governing TWA equations, and we derive the energy budget for this system in section \ref{sec:energy_budget}. We then describe our prognostic, one-equation model for the divergence of the EPFT in section \ref{sec:param}. Numerical simulations, described in section \ref{sec:numerical_simulations}, are then used to diagnose some of the terms in the model (section \ref{sec:diagnostics}) and then test it (section \ref{sec:results_test}). We end this paper with a discussion of our results, and some conclusions in section \ref{sec:discussion_conclusions}.

%%%%%%%%%%%%%%%%%%%%%%%%%%%%%%%%%%
\section{Governing Equations}
\label{s:govEqns}

In this section we present the thickness-weighted averaged (TWA) framework. 
This framework consists of the TWA Boussinesq equations for the residual mean flow, in which the effects of eddies appear as the divergence of the Eliassen-Palm Flux Tensor in the horizontal momentum equations, the equations for the fluctuating flow, and the equations for the mean and eddy kinetic and potential energy budgets.

\subsection{Residual-mean flow equations}

It is assumed that the Boussinesq equations have been averaged over microstructural length scales to produce a stably stratified field in which buoyancy decreases monotonically with depth. 
These equations are then weighted by thickness of buoyancy layers and ensemble-averaged \citep{deszoeke_bennett_1993, young_2012}.
Following  the notation in \cite{deszoeke_bennett_1993} and in \cite{young_2012}, the residual-mean flow equations in $z$-coordinates are
\begin{eqnarray}
\Dts{\hat{u}} - f \hat{v} + \frac{\partial p^\sharp }{\partial x} = \hat{R}_x - (\nabla \cdot \mathbf{E}) \cdot \mathbf{i} \; , \label{twaz_xmom} \\
\Dts{\hat{v}} + f\hat{u} + \frac{\partial p^\sharp }{\partial y} = \hat{R}_y - (\nabla \cdot \mathbf{E}) \cdot \mathbf{j} \; , \label{twaz_ymom}\\
\frac{\partial p^\sharp }{\partial z} = b^\sharp , \label{twaz_zmom} \\
\frac{\partial \hat u }{\partial x} + \frac{\partial \hat v }{\partial y} + \frac{\partial w^\sharp }{\partial z} = 0 , \label{twaz_continuity}\\
\Dts{b^\sharp} = \hat \varpi , \label{twaz_buoycons} \\
\Dts{\hat{c}} + \nabla \cdot \mathbf{J}^c = \hat \gamma , \label{twaz_tracers}
\end{eqnarray}
where $u$, $v$ and $w$ are zonal, meridional and vertical velocities, $p$ is pressure, $\hat R_x$ and $\hat R_y$ contain dissipation and forcing terms for the zonal and meridional momentum equations, $b$ is the buoyancy of a fluid parcel, $c$ is a tracer, and $x$, $y$, $z$ and $t$ represent the zonal, meridional, vertical locations, and time, respectively. 
Diabatic effects, or the velocity through buoyancy surfaces in buoyancy coordinates, are represented by $\hat \varpi$, and passive tracer diabatic effects by $\gamma$;
$f$ is the coriolis parameter;
$\mathbf{i}$ and $\mathbf{j}$ are the horizontal unit vectors in the zonal and meridional directions.
The caret or `hat' symbol $\hat{}\,\,$ is used to represent the thickness-weighted average of a quantity, e.g. 
\begin{equation}
\hat{u} = \frac{ \ol{u \sigma} }{ \ol{\sigma} },
\label{eq:twadefinition}
\end{equation}
where 
\begin{equation}
\ol{\sigma} = \left ( \ddz {b^\sharp} \right ) ^{-1}
\label{eq:sigma_definition}
\end{equation}
is the layer thickness per unit buoyancy. The overline symbol $^- \,$ represents an ensemble average at fixed $x, y, b$ and the sharp symbol $^\sharp\,\,$ indicates quantities that are dynamically consistent with the residual-mean flow without being averages themselves. The material derivative following the residual velocity in $z$-coordinates is given by
\begin{equation}
\Dts{ } = \ddt{} + \hat u \ddx{} + \hat v \ddy{} + w^\sharp \ddz{} \label{Dts},
\end{equation}
or, in buoyancy coordinates, by
\begin{equation}
\Dts{ } = \ddtb{} + \hat u \ddxb{} + \hat v \ddyb{} + \hat \varpi \ddb{} \label{Dts},
\end{equation}
where the tilde symbol $\; \tilde{} \;$ is used to indicate quantities at constant buoyancy.
The vector $\mathbf{J}^c$ is given by
\begin{equation}
\mathbf{J}^c = \widehat{u'' c''} \ol{\mathbf{e}}_1 + \widehat{v'' c''} \ol{ \mathbf{e}}_2 + \widehat{\varpi'' c''} \ol{ \mathbf{e}}_3
\end{equation}
and represents eddy fluxes of $c$.

The effects of eddies on the residual-mean flow appear in the horizontal momentum equations (\ref{twaz_xmom}-\ref{twaz_ymom}) as the divergence of the Eliassen-Palm flux tensor (EPFT) $\mathbf{E}$ \citep{maddison_marshall_2013},  given by
\begin{equation}
\mathbf{E} =
\begin{pmatrix}
  \vspace{0.2 cm}
  \widehat{ u'' u'' } +  \frac{1}{2 \overline \sigma} \ol{{\zeta'^2}} & \widehat{u''v''} & 0 \\
  \vspace{0.2 cm}
  \widehat{u''v''} & \widehat{v''v''} +  \frac{1}{2 \overline \sigma} \overline{{\zeta'^2}} & 0 \\
  \vspace{0.2 cm}
  \widehat{u'' \varpi''} + \frac{1}{\overline \sigma} \overline{{\zeta' m'_{\tilde x}}}  & \widehat{v'' \varpi''} +  \frac{1}{\overline \sigma} \overline{{\zeta' m'_{\tilde y}}} & 0
 \end{pmatrix} ,
\label{eq:fullEPFT}
\end{equation}
where $\zeta$ is the $z$-coordinate of a given buoyancy surface and $m = p/\rho_0 - b  \zeta$ is the Montgomery potential. Subscripts indicate derivatives with respect to $\tilde x$ and $\tilde y$ respectively.
The double prime represents deviations from TWA averaged quantities, e.g. 
\begin{equation}
u = \hat{u} + u'', 
\label{eq:doubleprime}
\end{equation}
and primed quantities represent deviations from the ensemble average, e.g. 
\begin{equation}
\zeta = \ol{\zeta} + \zeta '.
\label{eq:singleprime}
\end{equation}
The individual terms in the EPFT in (\ref{eq:fullEPFT}) represent eddy Reynolds stresses, or horizontal transfer of horizontal momentum ($\widehat{u''u''}$, $\widehat{u''v''}$ and $\widehat{v''v''}$), eddy potential energy ($-\frac{1}{2 \overline \sigma} \overline{{\zeta'^2}}$)  and eddy interfacial form drag or vertical transfer of horizontal momentum ($\frac{1}{\overline \sigma} \overline{{\zeta' m'_x}}$ and $\frac{1}{\overline \sigma} \overline{{\zeta' m'_y}}$), which is proportional to the eddy buoyancy fluxes (e.g. \cite{gent_mcwilliams_1990}).
The expression given in (\ref{eq:fullEPFT}) is only unique to within a rotational gauge term that is divergence-free and does not affect the dynamics of the flow \citep{maddison_marshall_2013}.
Thus, $( \nabla \cdot \mathbf{E} ) \cdot \mathbf{i}$ and $( \nabla \cdot \mathbf{E} ) \cdot \mathbf{j}$ in (\ref{twaz_xmom}-\ref{twaz_ymom}) are the zonal and meridional eddy forces on the residual mean flow, respectively, associated to eddy-mean flow interactions, which can also be linked to Ertel potential vorticity fluxes \citep[][among others]{young_2012, maddison_marshall_2013, saenz_etal_2015a}.

%We can decompose the EPFT into three terms,
%%
%\begin{equation}
%\mathbf{E} = \mathbf{E}_s + \mathbf{E}_d + \mathbf{E}_\varpi.
%\label{eq:fullEPFT_decomposition}
%\end{equation}
%%
%where
%%
%\begin{eqnarray}
%\mathbf{E}_s =
%\begin{pmatrix}
%  \vspace{0.2 cm}
%  \widehat{u''u''} +  \frac{1}{2 \overline \sigma} \overline{{\zeta'^2}} & \widehat{u''v''} & 0 \\
%  \vspace{0.2 cm}
%  \widehat{u''v''} & \widehat{v''v''} +  \frac{1}{2 \overline \sigma} \overline{{\zeta'^2}} & 0 \\
%  0 & 0 & 0
% \end{pmatrix}, 
%%
%\mathbf{E}_d =
%\begin{pmatrix}
%  \vspace{0.2 cm}
%  0 & 0 & 0 \\
%  \vspace{0.2 cm}
%  0 & 0 & 0 \\
%  \frac{1}{\overline \sigma} \overline{{\zeta' m'_{\tilde x}}}  & \frac{1}{\overline \sigma} \overline{{\zeta' m'_{\tilde y}}} & 0
% \end{pmatrix} \nonumber \\
%%
%\mathbf{E}_\varpi =
%\begin{pmatrix}
%  \vspace{0.2 cm}
%  0 & 0 & 0 \\
%  \vspace{0.2 cm}
%  0 & 0 & 0 \\
%  \widehat{u'' \varpi''} & \widehat{v'' \varpi''} & 0
% \end{pmatrix},
%\end{eqnarray}
%%
%For the time being, $\mathbf{E_s}$ and $\mathbf{E}_\varpi$ will be neglected and we will focus on parameterizing $\mathbf{E_d}$.
%However, the eddy parameterization framework in MPAS-O should be general enough to allow addition of parameterizations for the $\mathbf{E_s}$ and $\mathbf{E}_\varpi$ terms.

\subsection{Energy budget}
\label{sec:energy_budget}

Here we summarize the terms of the (mechanical) energy budget, defined as kinetic plus potential energy.
These are used to derive an expression for eddy energy, or the sum of eddy kinetic and potential energies.

We define mean kinetic energy as
\begin{equation}
\Km = \frac{\hat u ^2 + \hat v^2}{2} ,
\label{eq:def_mke}
\end{equation}
and eddy kinetic energy as
\begin{equation}
\Ke = \frac{ \widehat{ {u'' u''}} + \widehat{ {v'' v''} }}{2 } .
\label{eq:def_eke}
\end{equation}

After deriving the equation for the evolution of mean kinetic energy, we obtain a term for the work of the velocity field against the pressure gradient.
Then using the hydrostatic relation and the thickness equation, this term can be related to the material derivative of the potential energy of the residual-mean flow, thus arriving to an expression for potential energy for the TWA framework, namely
\begin{equation}
\Pm = \frac{1}{2 \ol \sigma} \ol{ \zeta}^2.
\label{eq:def_mpe}
\end{equation}
Using a process similar to the one used to derive the expression for $\Pm$, we use the pressure work terms in the eddy kinetic energy equation, and arrive to
\begin{equation}
\Pe = \frac{1}{2 \ol \sigma} \ol{\zeta' \zeta'}
\label{eq:def_epe}
\end{equation}

The equation governing the evolution of eddy energy $\Ee$, derived in \ref{a:eddyenergy}, is given by
\begin{equation}
\ddtb{\Ee} 
+ \mathbf{u^\sharp} \cdot \nabla \Ee 
=
  \hat u \, (\nabla \cdot \mathbf{E}) \cdot \mathbf{i}
+ \hat v (\nabla \cdot \mathbf{E}) \cdot \mathbf{j}
- \nabla \cdot \mathbf{T} 
+ \mathcal{D}_e,
\label{eqn:eddyEnergy}
\end{equation}
where $\mathbf{u^\sharp} = \hat u \mathbf{i} + \hat v \mathbf{j} + w^\sharp \mathbf{k}$. The terms on the left hand side of (\ref{eqn:eddyEnergy}) correspond to the time rate of change of eddy energy, transport of eddy energy by the mean flow, work from eddy-mean flow interactions through the EPFT in the zonal and meridional directions, and eddy energy transport by eddy fluxes $\mathbf{T}$, and on the right hand side we have eddy energy dissipation by adiabatic processes $\mathcal{D}_e$.

\section{Prognostic, One-Equation model for the divergence of the EPFT}
\label{sec:param}

As a first approach, or proof of concept to using the TWA formalism in equations (\ref{twaz_xmom} - \ref{twaz_tracers}) to parameterize eddy-mean flow interactions, \citet{saenz_etal_2015a} implemented the Gent-McWilliams eddy parameterization \citep[][GM90]{gent_mcwilliams_1990} in the TWA framework as a Coriolis force by the bolus velocity, ignoring Reynolds stress terms, eddy potential energy and diabatic terms. This implementation results in approximating the divergence of the EPFT as the vertical divergence of the eddy interfacial form drag terms, 
\begin{eqnarray}
(\nabla \cdot \mathbf{E}) \cdot \mathbf{i}= \frac{\partial }{\partial z} \left( \frac{1}{\overline \sigma} \overline{{\zeta' m'_{\tilde x}}} \right ) = f \frac{\ol{\sigma' v'}}{\ol{\sigma}} \equiv f v_*, \label{eq:thickness_fluxes_x} \\
(\nabla \cdot \mathbf{E}) \cdot \mathbf{j} = \frac{\partial }{\partial z} \left( \frac{1}{\overline \sigma} \overline{{\zeta' m'_{\tilde y}}} \right ) = -f \frac{\ol{\sigma' u'}}{\ol{\sigma}} \equiv - f u_* \label{eq:thickness_fluxes_y}  .
\end{eqnarray}
where the horizontal bolus velocity vector in cartesian coordinates $\mathbf{u}_* = u_* \mathbf{i} + v_* \mathbf{ j} $ can be written as
\begin{equation}
\mathbf{u}_* = - \ddz{} \left( \kappa \, \mathbf{S} \right ) \label{GMparameterization} .
\end{equation}
Eddy diffusivity $\kappa$ is set to a constant.

\subsection{Eddies as viscous forces in the horizontal momentum equations}

An alternative to the GM90 parameterization, and perhaps a better choice \citep{greatbatch_lamb_1990, ferreira_marshall_2006}, is to represent eddies as viscous terms in the horizontal momentum equation, in which
\begin{eqnarray}
\frac{\overline{{\zeta' m'_{\tilde x}}} } {\ol \sigma} = - \mu \ddz{ \hat{u} },  \label{eq:formdrag_paramx} \\
\frac{\overline{{\zeta' m'_{\tilde y}}} } {\ol \sigma} = - \mu \ddz{ \hat{v} }. \label{eq:formdrag_paramy}
\end{eqnarray}
The eddy viscocity $\mu$ can be related to $\kappa$ via $\mu = \kappa f^2 \ol N^{-2}$ \citep{greatbatch_lamb_1990, saenz_etal_2015a}. We take this approach, and in this section we propose a prognostic eddy parameterization for the vertical viscosity in equations (\ref{eq:formdrag_paramx}-\ref{eq:formdrag_paramy}) in which we use mixing length theory to relate $\mu$ to a prognostic, vertically integrated eddy energy.
We assume that the only forces are from the interfacial form drag terms in the EPFT, while the forces from diabatic, sheer and eddy potential energy terms are zero, similar to the assumptions made in GM90.
We represent the dominant baroclinic mode through a simplified model and use it to represent the vertical structure of $\mu$ as a function of the residual mean flow. 
We design the framework to be independent of the specification of the model for the baroclinic mode, allowing, in principle, the inclusion of more realistic models for baroclinic mode decomposition and the inclusion of more than one mode in our model.

We start by defining the vertically integrated eddy energy, $\IEe$, as a function of $x,y,t$, where
\begin{equation}
\langle \cdot \rangle_z(x,y,t) = \int_{z_b(x,y)} ^{\eta(x,y,t)} \cdot \; dz,
\label{eq:vert_integral_eddy_energy}
\end{equation}
$\eta(x,y,t)$ is the sea surface height at a given location and time, and $z_b(x,y)$ is the bottom depth.
Ignoring the eddy energy transport by the mean flow and by the eddy fluxes, the second and last terms on  the left hand side of (\ref{eqn:eddyEnergy}), and integrating (\ref{eqn:eddyEnergy}) vertically we obtain
\begin{equation}
\ddtb{\IEe} 
=
   \langle \hat u \, (\nabla \cdot \mathbf{E}) \cdot \mathbf{i} \rangle_z
+ \langle \hat v (\nabla \cdot \mathbf{E}) \cdot \mathbf{j} \rangle_z
+ \langle \mathcal{D}_e \rangle_z \;.
\label{eqn:VertIntEddyEnergy}
\end{equation}
We use the above to construct a prognostic representation of the eddy energy in the flow.

We represent the eddy interfacial form drag terms in the bottom row of the EPFT (\ref{eq:fullEPFT}) using equations (\ref{eq:formdrag_paramx}-\ref{eq:formdrag_paramy}).
This way, the forces from eddy mean flow interactions in the horizontal momentum equations (\ref{twaz_xmom}-\ref{twaz_ymom}) and in the eddy energy equation (\ref{eqn:eddyEnergy}) and (\ref{eqn:VertIntEddyEnergy}) become
\begin{eqnarray}
(\nabla \cdot \mathbf{E}) \cdot \mathbf{i} = - \ddz{} \left ( \mu_e \ddz{ \langle \hat{u} \rangle_{t_e} } \right ), \label{eq:paramForceX} \\
(\nabla \cdot \mathbf{E}) \cdot \mathbf{j} = - \ddz{} \left ( \mu_e \ddz{ \langle \hat{v} \rangle_{t_e}} \right ). \label{eq:paramForceY}
\end{eqnarray}
In the above, $\langle \; \rangle_{t_e}$ is a time filter used to average the velocity at eddy time-scales $t_e$.
We use mixing length theory to estimate the diffusivity,
% We originally decided to use eddy energy and not eddy kinetic energy, based on:
% Marshall et al 2012 A Framework for Parameterizing Eddy Potential Vorticity Fluxes JPO.
%
\begin{equation}
\kappa_e = \sqrt{\Ee} \, \ell_m
\end{equation}
and, following \cite{greatbatch_lamb_1990} and \cite{saenz_etal_2015a}, we use $\mu_e = K f^2 \ol N^{-2}$ to relate the viscosity to the diffusivity, resulting in
\begin{equation}
\mu_e = \frac{f^2}{N^2} \sqrt{\Ee} \, \ell_{m},
\end{equation}
where $\ell_m$ is the mixing length and $\Ee$ is the eddy energy. 
This relation is based on the work of \cite{green_1970}, \cite{stone_1972}, and has been used in \cite{eden_greatbatch_2008}, \cite{marshall_adcroft_2010}, among others.
The mixing length can be represented as a function of the eddy length scale $L_e$. 
A simple model for the mixing length is
\begin{equation}
\ell_m = \alpha_{eff} \, L_e
\end{equation}
where $\alpha_{eff} \equiv \alpha_{eff}(z)$ is an efficiency factor, which is sometimes `absorbed' into $L_e$, that takes into account the vertical structure of the mixing length, e.g. as a function of eddy suppression \citep{ferrari_nikurashin_2010, smith_2007, smith_marshall_2009, bates_etal_2014}.
Here we will limit ourselves to the simplest mixing length theory model, and we use 
\begin{equation}
\mu_e = \frac{f^2}{N^2} \sqrt{\Ee} \,  L_e \;.
\label{eq:eddy_viscosity}
\end{equation}
We relate the eddy energy $\Ee$ to the prognostic, vertically integrated eddy energy $\IEe$ through a model for normal mode decomposition of each water column,
\begin{equation}
\Ee(x,y,z,t) = \phi(x,y,z,t) ^2 \IEe (x,y,t) \; .
\end{equation}

As is common in ocean models, we ignore momentum eddy fluxes in and out of the ocean.
We implement this constraint by enforcing the boundary conditions at the boundaries of the ocean as
\begin{equation}
\mu(x,y,\eta,t) = \mu(x,y,z_b,t) = 0 \; .
\end{equation}
%

%%%%%%%%%%%%%%%%%%%%%%%%%%%%%%%%%%
\section{Specific Formulation and Algorithm}

In this section we present specific formulations for the time filters and the baroclinic mode decomposition model that we used for this paper.

The dissipation term in equation (\ref{eqn:VertIntEddyEnergy}) is represented using the following simple algebraic model, 
\begin{equation}
\langle \mathcal{D}_e \rangle_z = \gamma_e \IEe \;,
\label{eqn:VertIntEddyEnergyDissipation}
\end{equation}
where $\gamma_e$ is an inverse eddy dissipation time-scale.

The time filter $\langle \; \rangle_{t_i}$ is calculated using an impulse model filter \citep{smith_book_1997}.
For a quantity $\psi^n = \psi(x,y,z,t_n)$ at an $n$th timestep $t_n$, the filtered quantity $\langle \psi \rangle_{t_i}$ is updated using 
\begin{equation}
\langle \psi \rangle_{t_i} = \langle \psi \rangle^n_{t_i} + \frac{\Delta t_i}{t_i} (\psi^{n} - \langle \psi \rangle_{t_i}) \\
\label{eq:time_filter}
\end{equation}
where $\Delta t_i$ is the frequency of the sampling, and $t_i$ is the span of the moving average.

To test our formulation, we first implement a simplified model for the baroclinic mode decomposition in a water column.
We assume that the first baroclinic mode is the most energetic and dominates the dynamics of the flow.
It has been noted in previous studies that this is not the case, in different parts of the ocean other modes are more energetic, and sometimes a combination of modes contain most of the eddy energy \citep{smith_2007, smith_marshall_2009}.
However, for the purpose of this paper we implement a simplified model, but in future work this model can be replaced by more realistic ones as the ones.

We express the first baroclinic mode as 
\begin{equation}
\phi(x,y,z,t) = - A(x,y,t) \exp{} \left(\frac{z-\eta(x,y,t)}{H_e} \right ).
\label{eq:baroclinic_mode}
\end{equation}
The $e$-folding depth $H_e(x,y,t)$ is set as
\begin{equation}
H_e = \min (\psi_1 H(x,y), H_{1}),
\end{equation}
where $\psi_1$ is a parameter indicating a fraction of the column depth, and $H_1$ is a parameter representing the minimum $e$-folding depth.
The amplitude of the eigenvector $A(x,y,t)$ is calculated by numerically enforcing
\begin{equation}
\int_{-H(x,y)}^{\eta(x,y,t)} \left[ \phi(x,y,z,t) \right ]^2 dz = 1 \;.
\label{eqn:phisq}
\end{equation}

The eddy length scale in equation (\ref{eq:eddy_viscosity}) is calculated using the Wentzel-Kramers-Brillouin approximation of the Rossby Radius of deformation \citep{chelton_etal_1998},
\begin{equation}
L_e = \frac{1}{| f |} \int_{-H(x,y,t)}^{0} \langle N \rangle_{t_c} \;dz,
\label{eq:eddyLengthScale}
\end{equation}
where the time filter $\langle \; \rangle_{t_c}$ is used to average the buoyancy frequency over a climatological time scale $t_c$.

\subsection{Algorithm}
\label{s:algorithm}

Given the state of the flow at a time-step $n$, before performing timestep $n+1$ in the dynamical core, and the viscosity $\mu_n$, the algorithm proceeds as follows.

\begin{enumerate}

\item Update the time filtered quantities given quantities from the previous time-step, using (\ref{eq:time_filter}); i.e. update $\langle{u}\rangle_{t_e}$ and $\langle{N^2}\rangle_{t_c}$.
\label{step:1}

\item Calculate vertical derivative
\begin{equation}
\ddz{ }\langle{u}\rangle_{t_e}.
\end{equation}

\item Calculate vertical flux of horizontal momentum
\begin{equation}
\mu \ddz{ }\langle{u}\rangle_{t_e}.
\end{equation}

\item Calculate the force, or the (negative) time tendency in momentum (\ref{twaz_xmom}), as the vertical divergence of the vertical flux of horizontal momentum,
\begin{equation}
(\nabla \cdot \mathbf{E}) \cdot \mathbf{i} = - \ddz{ } \left( \mu_{n} \ddz{ }\ol u_{n} \right) .
\label{eq:twaz_xmom_tend}
\end{equation}

\item Calculate the time tendency of the eddy energy
\begin{equation}
\hat u \, (\nabla \cdot \mathbf{E}) \cdot \mathbf{i} \;.
\label{eq:eddyEnergy_tend}
\end{equation}

\item Using (\ref{eq:vert_integral_eddy_energy}), sum over the column to get the eddy energy tendency over the column \\
\begin{equation}
\langle \hat u \, (\nabla \cdot \mathbf{E}) \cdot \mathbf{i} \rangle_z \;.
\label{eq:vertIntEddyEnergy_tend}
\end{equation}

\item Integrate the vertically integrated eddy energy (\ref{eqn:VertIntEddyEnergy}) forward in time for one time-step to obtain $\langle \Ee \rangle^{n+1}_z$, using the tendency calculated in (\ref{eq:vertIntEddyEnergy_tend}). 
Here, we use the Euler method, representing the time tendency from the transfer of energy from the mean flow (first and second terms on the right hand side of equation (\ref{eqn:VertIntEddyEnergy})) explicitly, and the dissipation in equation (\ref{eqn:VertIntEddyEnergyDissipation}) implicitly for stability.

\item Solve the eigenvalue problem, equation (\ref{eq:baroclinic_mode}), for $\phi(x,y,z,t)$.

\item Calculate the eddy length scale (\ref{eq:eddyLengthScale}), namely
\begin{equation}
L = \frac{1}{| f |} \int_z \sqrt{\langle N^2 \rangle_{t_c} } dz \; .
\end{equation}

\item Calculate the eddy energy along the water column, by projecting the vertically integrated eddy energy onto the water column using the $\phi(x,y,z,t)$,
\begin{equation}
\Ee(x,y,z) = \langle \Ee \rangle^{n+1}_z  \,\,\, \phi^2(x,y,z) \; .
\end{equation}

\item Update the eddy viscosity 
\begin{equation}
\mu_{n+1}(x,y,z) = \frac{f^2}{\langle N^2 \rangle_{t_c} } \sqrt{ \Ee(x,y,z) } \, L \, \alpha_{eff}.
\label{eq:visc_in_algorithm}
\end{equation}

\item Advance the dynamical code forward one timestep, in which the state of the flow us updated to timestep $n+1$ and the momentum equation (\ref{twaz_xmom}) is updated with the time tendency from eddy forces given by (\ref{eq:twaz_xmom_tend}) calculated above.

\item If the dynamical core will perform another cycle, go to step \ref{step:1}.

\end{enumerate}

The above algorithm is designed so that the vertically integrated eddy energy is conserved by construction, since the time tendency of momentum from the parameterized eddies (\ref{eq:twaz_xmom_tend}) and the time tendency of vertically integrated eddy energy by the parameterized eddies (\ref{eq:vertIntEddyEnergy_tend}) are consistent with the derivation of (\ref{eqn:VertIntEddyEnergy}).

\section{Numerical Simulations}
\label{sec:numerical_simulations}

We implement the one-equation prognostic model and the algorithm presented in section (\ref{s:algorithm}), into the ocean component of the Model for Prediction Across Scales \citep[MPAS-O;][]{ringler_etal_2013}.
The numerical scheme in MPAS-O solves the Boussinesq equations (\ref{twaz_ymom}-\ref{twaz_tracers}) using a mimetic, finite-volume discretization on a horizontal mesh consisting of Voronoi tesselations \citep{thuburn_etal_2009, ringler_etal_2010}, and an arbitrary Lagrangian-Eulerian (ALE) vertical coordinate \citep{petersen_etal_2014}.
Equations for scalars are expressed in flux form and are advected using a monotone transport algorithm \citep{skamarock_gassmann_2011}.

To test our implementation, we use the same configuration as in \cite{ringler_etal_2017} to setup and simulate the flow in a zonal idealized Southern Ocean (ZISO) configuration, in which 5 km resolution was used. 
Tests with the configuration in \cite{ringler_etal_2017} indicate that using 5km resolution and 10 km resolution, the flows are very similar.
So, to isolate effects of our model from the effects of changes in resolution, we choose a 10 km grid size.
We chose 10 km resolution and not 5 km in order to make our test case run quicker for the purpose of testing and debugging.
We use a horizontal mesh representing a 40 km $\times$ 2000 km region with a nominal resolution of 10 km, consisting of 920 tessellated hexagons arranged on a 4 $\times$ 230 mesh.
The vertical mesh has 100 layers spanning a depth of 2500 m in the main channel, with resolution varying between 0.63 m at the surface, and 92 m at the bottom.
About half of the layers in the vertical mesh are located in the upper 250 m of the water column.
The domain is periodic in the zonal direction $x$, and in the meridional direction $y$ it is bound by walls with no-slip boundary conditions.

The bathymetry consists of, from the south to the north, a continental shelf with a depth of $H_s = $500 m and a width of about 300 km, followed by a shelf break region of about 300 km where the depth increases towards the north, down to the bottom of the main channel where $z = -H = -2500$ m. 
This is achieved using the following function, as in \cite{ringler_etal_2017},
\begin{equation}
h(y) = H_s + \frac{1}{2}(H-H_s) \left [ 1 + \tanh \left( \frac{y-Y_s}{W_s} \right ) \right ],
\end{equation}
where $y=Y_s = $ 250 km and $y=W_s = $ 500 km are the center positions of the shelf and the shelf break, respectively.

We use an interior restoring model to represent the production of deep bottom water over the shelf at $y=0$ km, and to model the portion of the meridional overturning circulation that lies to the north of the channel by transforming north-flowing deep bottom waters into south-flowing mid-depth waters at the northern boundary, $y = 2000$ km.
The interior restoring of temperature $T_i$ is set using a temperature tendency at the boundary at a prescribed time-scale $\tau_i$,
\begin{equation}
\frac{d T_b}{d t} = - \frac{T - T_i}{\tau_i},
\end{equation}
where the interior restoring temperature $T_i$ is prescribed as
\begin{equation}
\label{equation:interiorRestoring}
T_i\left(y,\,z\right)=T_{r}\exp\left(\frac{z}{z_{e}}\right),
\end{equation}
and
\begin{equation}
\label{equation:interiorRestoringNorthWall}
\tau_i=\frac{1}{\tau_{T_{r}}}\left(e^{-y'/L_{e}}\right),
\end{equation}
\begin{equation}
  y' = \begin{cases}
    y & \textrm{for}\;\;y=0\;\textrm{km} \\
    L_y-y & \textrm{for}\;\;y=2000\;\textrm{km},
   \end{cases}
\end{equation}
and $z_e$=1000~m, $\tau_{T_{r}}=30$~days, $L_e=80$~km, and $L_y=2000$~km.

Surface wind forcing is specified using the wind stress
\begin{equation}
\tau(y) = \left \{ 
  \begin{array}{ll}
    -\tau_f \sin \left ( \pi \frac{L_s - y}{L_s} \right ) ^2 & : y \le L_s\\
    0 							& : L_s > y > L_s - W_{sf},  \\
    \tau_0 \sin \left ( \pi \frac{y-L_s}{L_y - L_s} \right ) ^2 & : y \le L_s
  \end{array}
  \right .
\end{equation}
where $\tau_0 = $ 0.2 N/m$^2$ is the peak eastward wind stress over the main channel, and $\tau_f = $ 0.05 N/m$^2$ is the peak westward wind stress over the shelf region.

Buoyancy forcing is achieved by restoring temperature at the surface to
\begin{equation}
T_r(y) = T_m + T_a \tanh \left ( 2 \frac{y - L_y/2}{L_y/2} \right ) + T_b \frac{y - L_y/2}{L_y/2},
\end{equation}
using a `piston' temperature flux of the form
\begin{equation}
Q_T = -p (T - T_r), 
\end{equation}
where $T_m = $ 3.0$^\circ$C, $T_a = $ 1.0$^\circ$C, $T_b = 2.0$$^\circ$C, and $p = $ 1.0$\times$10$^{-5}$ m/s is the piston velocity.

We use a linear equation of state,
\begin{equation}
\label{equation:linearEOS}
\rho=\rho_{ref}-\alpha\left(T-T_{ref}\right),
\end{equation}
where $\rho_{ref}=1025.0$ kg/m$^{3}$, $\alpha=0.255$ kg/m$^{3}/^\circ$C, and $T_{ref}=19.0 \,^\circ$C.

To stabilize the the simulations, we use a biharmonic viscosity $\nu_h = 6.25\times10^9$ m$^4/s$ for the divergent part of the flow, which is scaled by a factor of 10 for the rotational part of the flow.
We parameterize the bottom boundary layer stress using a quadratic drag with a coefficient of $3.0\times10^{-3}$.
We use the $K$-profile parameterization \citep[KPP,][]{large_etal_1994}, with background diffusivity and viscosity of $5\times10^{-6}$ and $1\times10^{-4}$ m$^2$/s, respectively.

\section{Diagnostic Evaluation of the Two-Equation Model in an Eddy Resolving Simulation}
\label{sec:diagnostics}

\begin{figure}[t]
\centering
\noindent\includegraphics[width=0.8\textheight,angle=0]{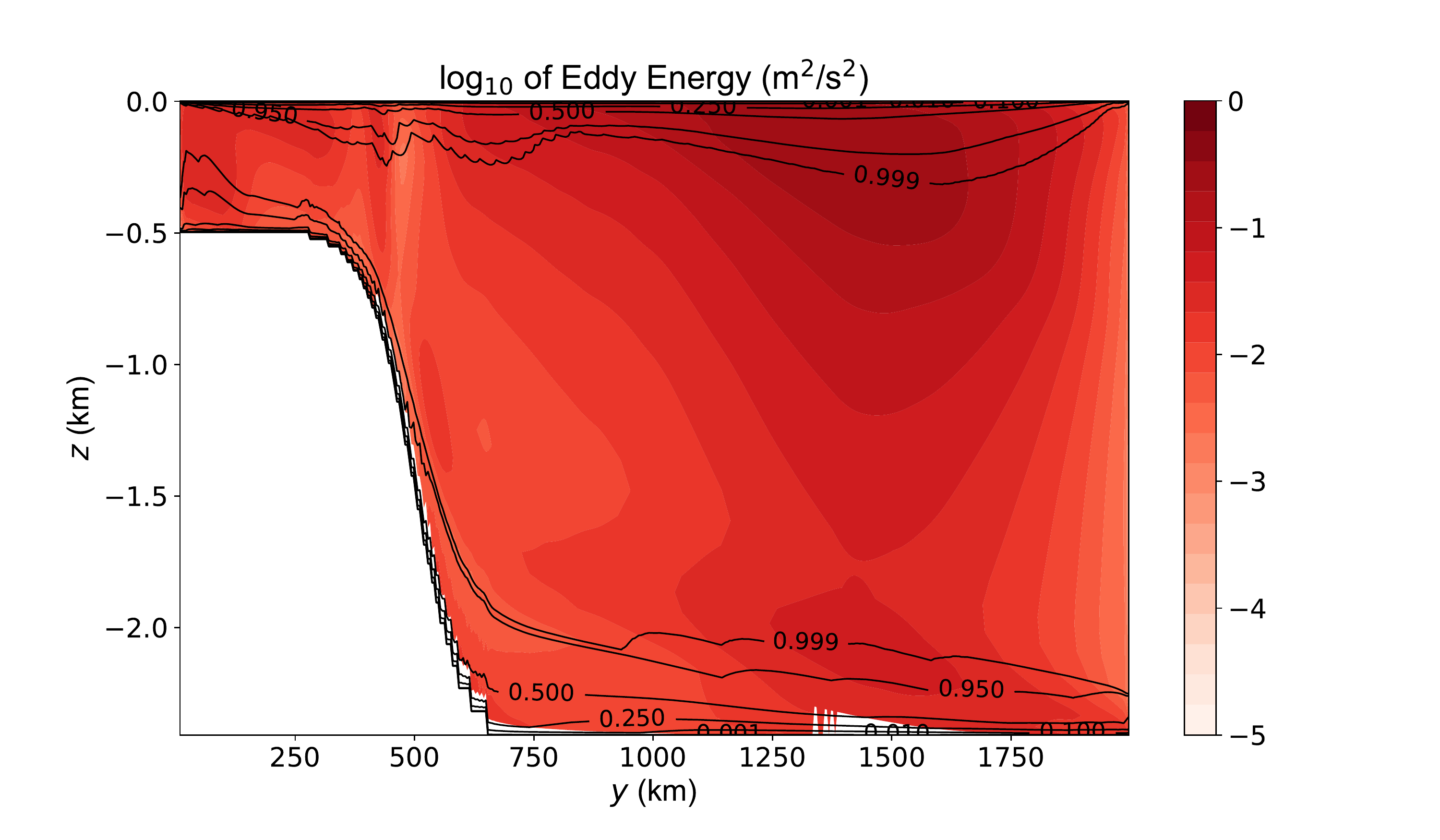}
\caption{Zonally averaged eddy energy $E_{ii}$ in the high resolution simulation simulation, plotted in log$_{10}$ scale (pseudo-color). Black contours show the probability of existence of 0.001, 0.01, 0.10, 0.25, 0.50, 0.95, 0.999.}
\label{fig:eddy_energy}
\end{figure}

In this section, we use our model in diagnostic mode by evaluating terms in the model using the high resolution, eddy resolving zonal idealized southern ocean (hrZISO) simulation in \cite{ringler_etal_2017}.
The hrZISO is characterized by a surface intensified mean flow towards the East ($\hat u > 0$) in the main channel.
We diagnose the eddy viscosity using the definition $\mathbf{E}$ in (\ref{eq:fullEPFT}), assuming adiabatic flow in the interior, and rewriting (\ref{eq:formdrag_paramx}) to obtain
\begin{equation}
\mu = \frac{E_{31}}{\frac{\partial u }{ \partial z}},
\label{eq:diagnosed_viscosity}
\end{equation}
where we have expressed the EPFT in index notation.
The eddy form drag $E_{31}$ and the vertical gradient of the zonally averaged, zonal residual mean velocity are calculated in buoyancy coordinates, as described in hrZISO.
In addition, we calculate the viscosity using the model in (\ref{eq:eddy_viscosity}), where the eddy energy is calculated using the trace of the EPFT $\Ee = E_{ii}$, and $E_ij$, $f$, $N^2$ or obtained from hrZISO.
Figure \ref{fig:eddy_energy} shows the eddy energy, along with the probability of existence $\ol \phi$, i.e the frequency at which a water mass of a given buoyancy class exists in the interior of the hrZISO \citep{ringler_etal_2017}.
We will define regions where $\ol \phi <1$ as ventilated regions, and those where $\ol \phi =1$ as interior regions.
The eddy energy is concentrated towards the surface of the circumpolar current in the main channel, a large portion of which corresponds to the ventilated region of the water column, as indicated by the depths in which the probability of occurrence is less than one.

\begin{figure}[t]
\centering
\noindent\includegraphics[width=0.8\textheight,angle=0]{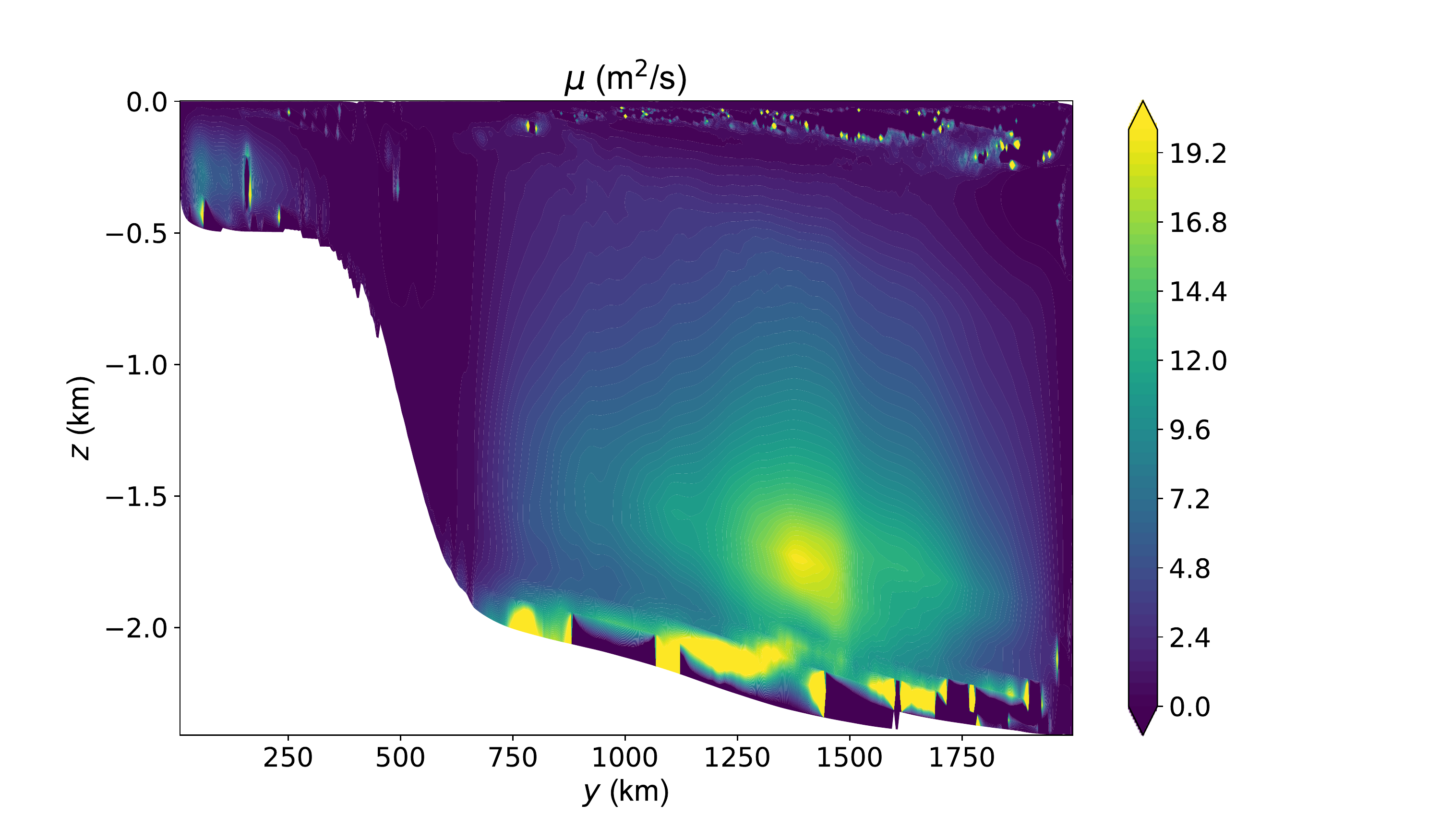} \\
\noindent\includegraphics[width=0.8\textheight,angle=0]{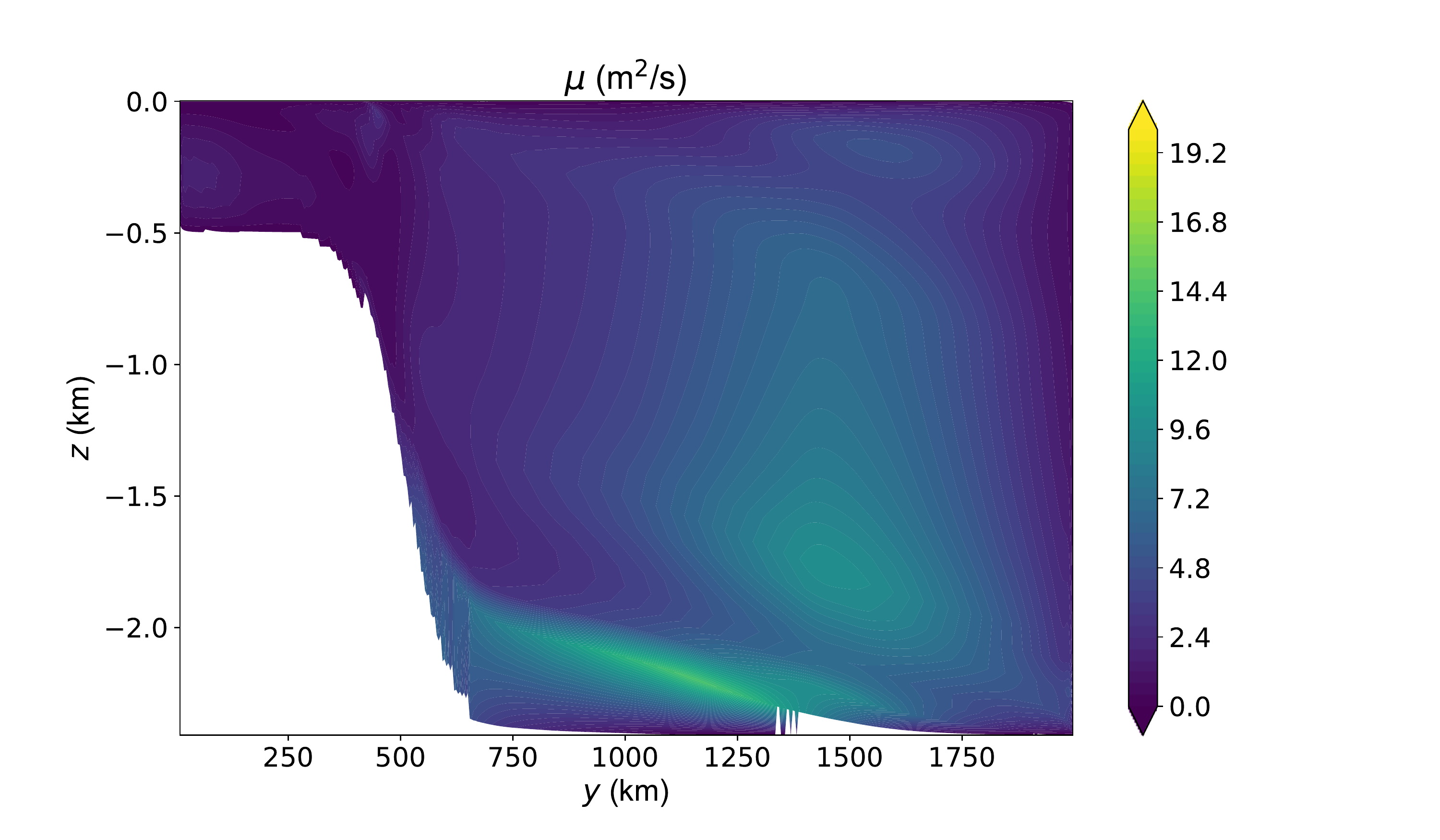}
\caption{Eddy viscosity diagnosed in the high resolution simulation hrZISO (top), and calculated with the proposed model, using terms diagnosed from the high resolution simulation hrZISO (bottom).}
\label{fig:visc_hrZISO}
\end{figure}

The diagnosed and modeled eddy viscosities calculated as described above, using data from hrZISO, are shown in figure \ref{fig:visc_hrZISO}.
There is good qualitative agreement, in that the viscosity in the interior of the flow is large inside the circumpolar current of the hrZISO, and within this current, it is largest at a depth of about 1.7 km.
Within the top and bottom 'ventilated' regions (where the probability of occurrence is less than 1 in figure \ref{fig:eddy_energy}), the diagnosed viscosity is noisy, and the modeled viscosity is large.
However, we do not have a model for the dynamics that occur in the ventilated region, and this remains to be investigated further.

\begin{figure}[t]
\centering
\noindent\includegraphics[width=0.8\textheight,angle=0]{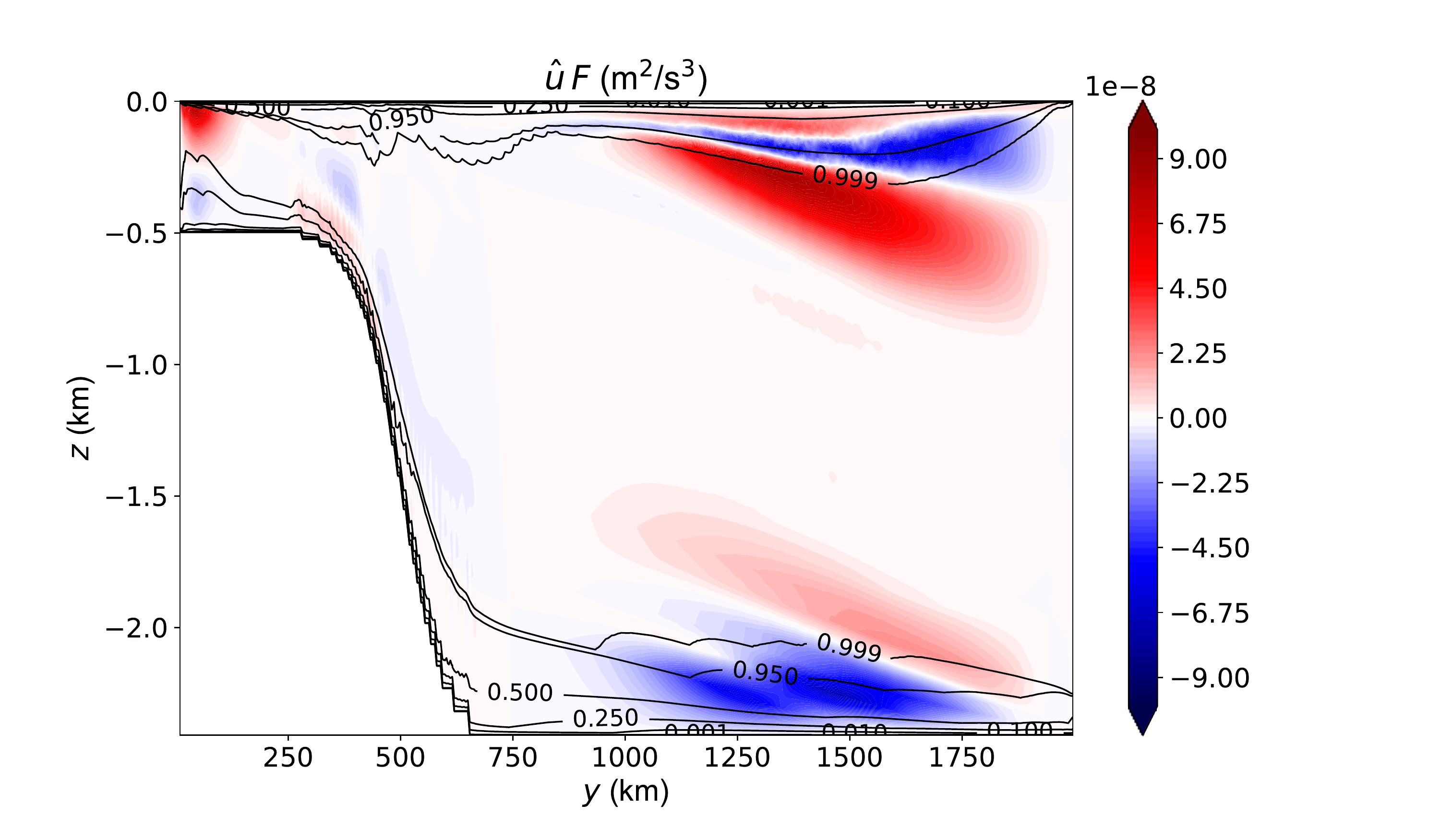} \\
\noindent\includegraphics[width=0.8\textheight,angle=0]{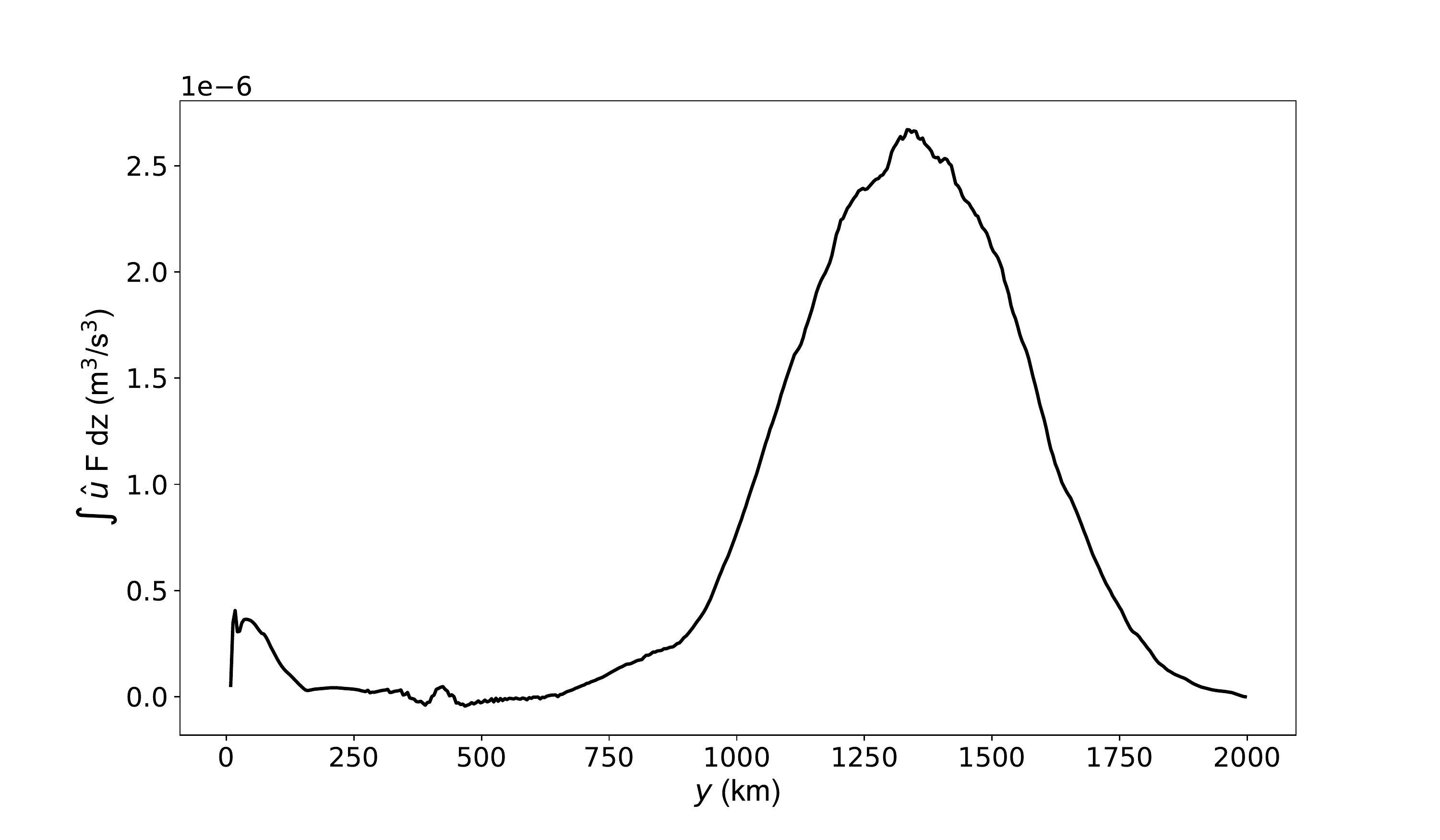}
\caption{Work by mean flow in the eddies diagnosed from the high resolution simulation, hrZISO. Black contours show the probability of existence of 0.001, 0.01, 0.10, 0.25, 0.50, 0.95, 0.999 (top); and vertically integrated work by the mean flow on the eddies (bottom).}
\label{fig:work_diag}
\end{figure}

The eddy forces in this flow are dominated by the eddy form drag in the zonal direction, and are given by $F \equiv (\nabla \cdot \mathbf{E}) \cdot \mathbf{i} = \partial E_{31} / \partial z$.
The rate of work of the mean flow on the eddies $\hat u \, F$ is plotted in figure \ref{fig:work_diag}.
By equation (\ref{eq:mke}), in regions where $\hat u \, F$ is positive, eddies are extracting energy from the mean flow, or the mean flow is doing work on the eddy field.
The mean zonal flow in the interior region of the main channel ($\ol \phi =1$, top of figure \ref{fig:work_diag}) is extracting energy from the eddies, and the vertically integrated work is positive (bottom of figure \ref{fig:work_diag}) as the eddies transfer momentum from the top of the channel, where it is input by wind stress, towards the bottom, where it is absorbed by bottom drag, as discussed in more detail in \cite{ringler_etal_2017}.

\section{Result from Simulations using the Two-Equation Model}
\label{sec:results_test}

To get the simulation to be stable, we had to limit the value of $\langle N^2_{tc} \rangle$ in equation (\ref{eq:visc_in_algorithm}).
We have tried a number of approaches, and we currently calculate $\langle N^2_{tc} \rangle$ as $N^2$ vertically averaged within a water column at a given horizontal location $(x,y)$.
With this regularization, we are able to integrate forward in time, but the flow we obtain does not well represent the average flow observed in the hrZISO.

To test the two-equation model in prognostic mode in a numerical simulation of the ZISO, we spin up the simulation from rest, in two steps.
First we simulate the Eulerian mean flow using the GM parameterization for 28 years, at which point we have a steady state.
Then we change to thickness-weighted coordinates by transforming the prognostic velocity to residual-mean velocity, and we run for another 8 years.
We refer to this spun up simulation as the parameterized ZISO (pZISO).

The flow is zonally uniform in the main channel, so when we show cross section plots, we will be showing $z,y$ slices at fixed $x$.

\begin{figure}[t]
\centering
\noindent\includegraphics[width=0.8\textheight,angle=0]{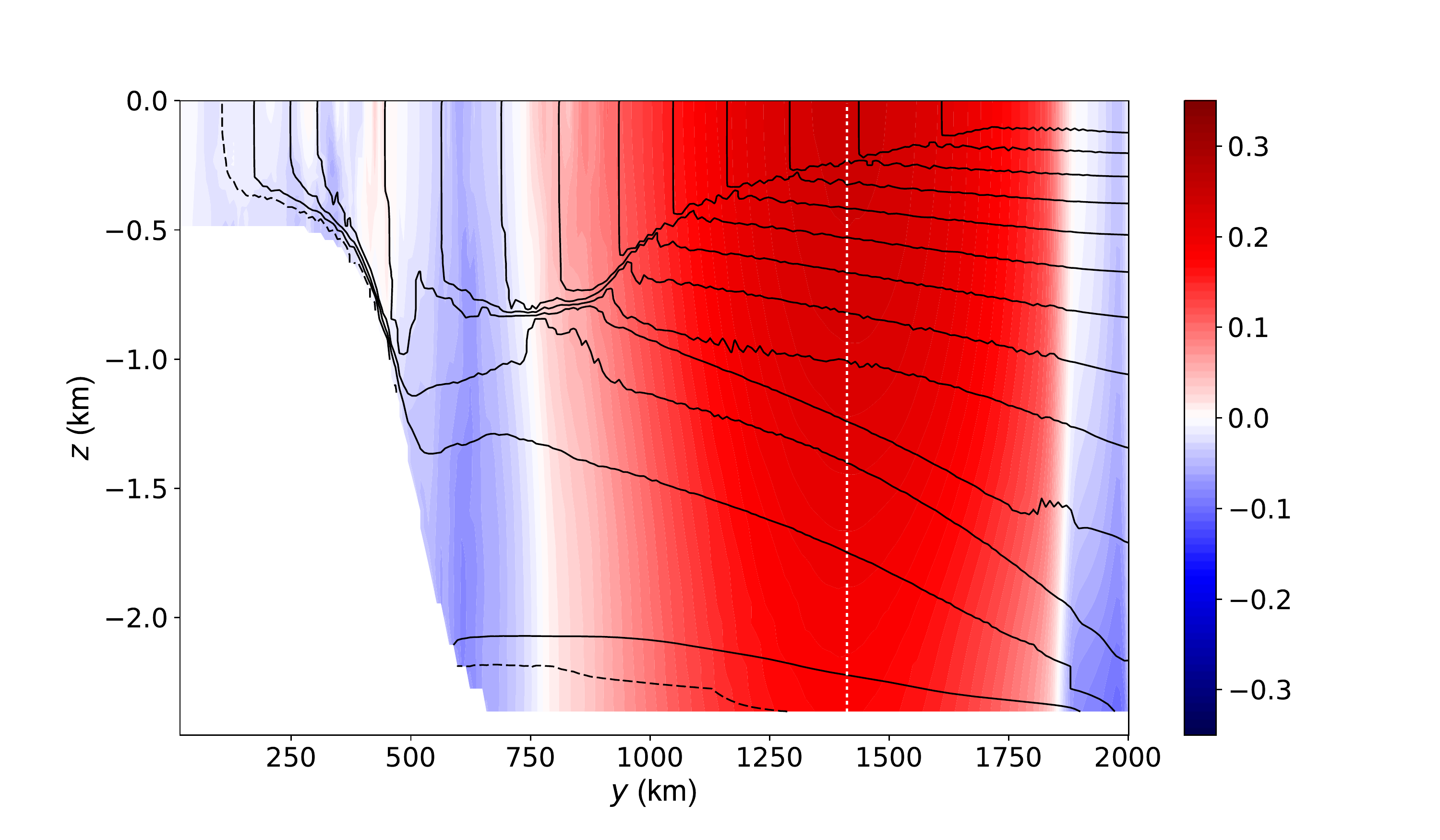}
\caption{Residual mean zonal velocity (color, in m/s$^2$) and temperature contours ($^{\circ}$C), starting at -0.5$^{\circ}$ (dashed line) with 0.5$^{\circ}$ intervals.}
\label{fig:par_utwa}
\end{figure}

Figure \ref{fig:par_utwa} shows a meridional cross-section of residual mean zonal velocity $\hat u$, along with temperature contours in pZISO.
The velocity obtained with pZISO is very similar to the velocity simulated in hrZISO, with an surface intensified, eastward zonal mean flow in the main channel that peaks at about 0.3 m/s at the surface. 
There is a westward current centered at about $y=740$ km, over the deepest end of the continental slope, while in the hrZISO this current is distributed of the entire range of the continental slope, peaking at around $y=500$ km.
The temperature contours in figure \ref{fig:par_utwa} differ from the ones observed in the hrZISO, in that the former have a substantially deeper mixed layer than the latter.

\begin{figure}[t]
\centering
\noindent\includegraphics[width=0.8\textheight,angle=0]{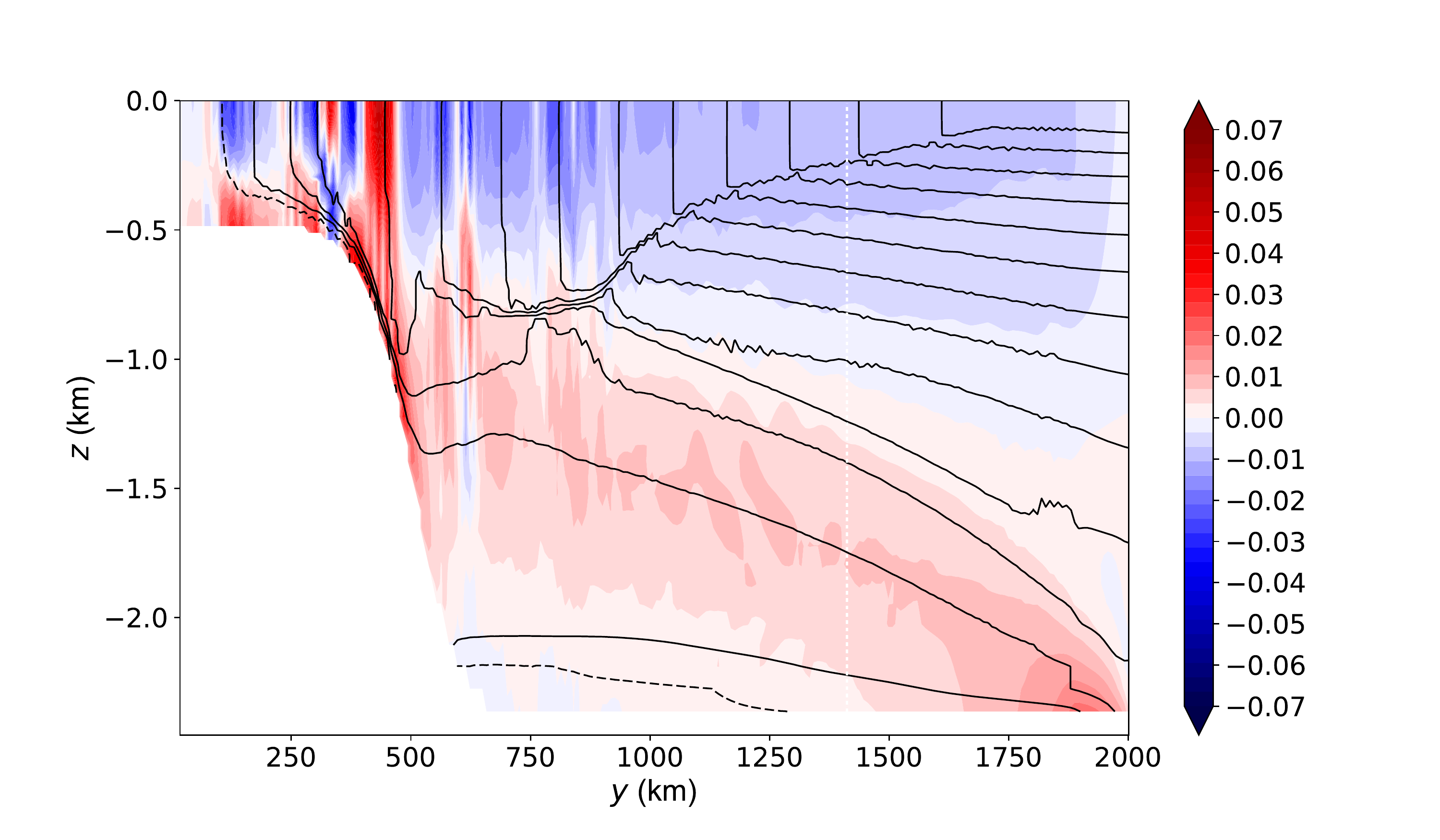}
\caption{Residual mean meridional velocity (color, in m/s$^2$) and temperature contours ($^{\circ}$C), starting at -0.5$^{\circ}$ (dashed line) with 0.5$^{\circ}$ intervals.}
\label{fig:par_vtwa}
\end{figure}

The residual mean meridional velocity is plotted in figure \ref{fig:par_vtwa}.
The simulation using our parameterization does not capture the detailed north/south meridional within the surface ventilated region, as we do not attempt to model it.
However, within the interior region of the pZISO main channel, the flow is southward up to a depth of about 1.3 km, and below this depth it is towards the north, with magnitudes that are within reasonable agreement between the parameterized and high resolution ZISO simulations.

\begin{figure}[t]
\centering
\noindent\includegraphics[width=0.8\textheight,angle=0]{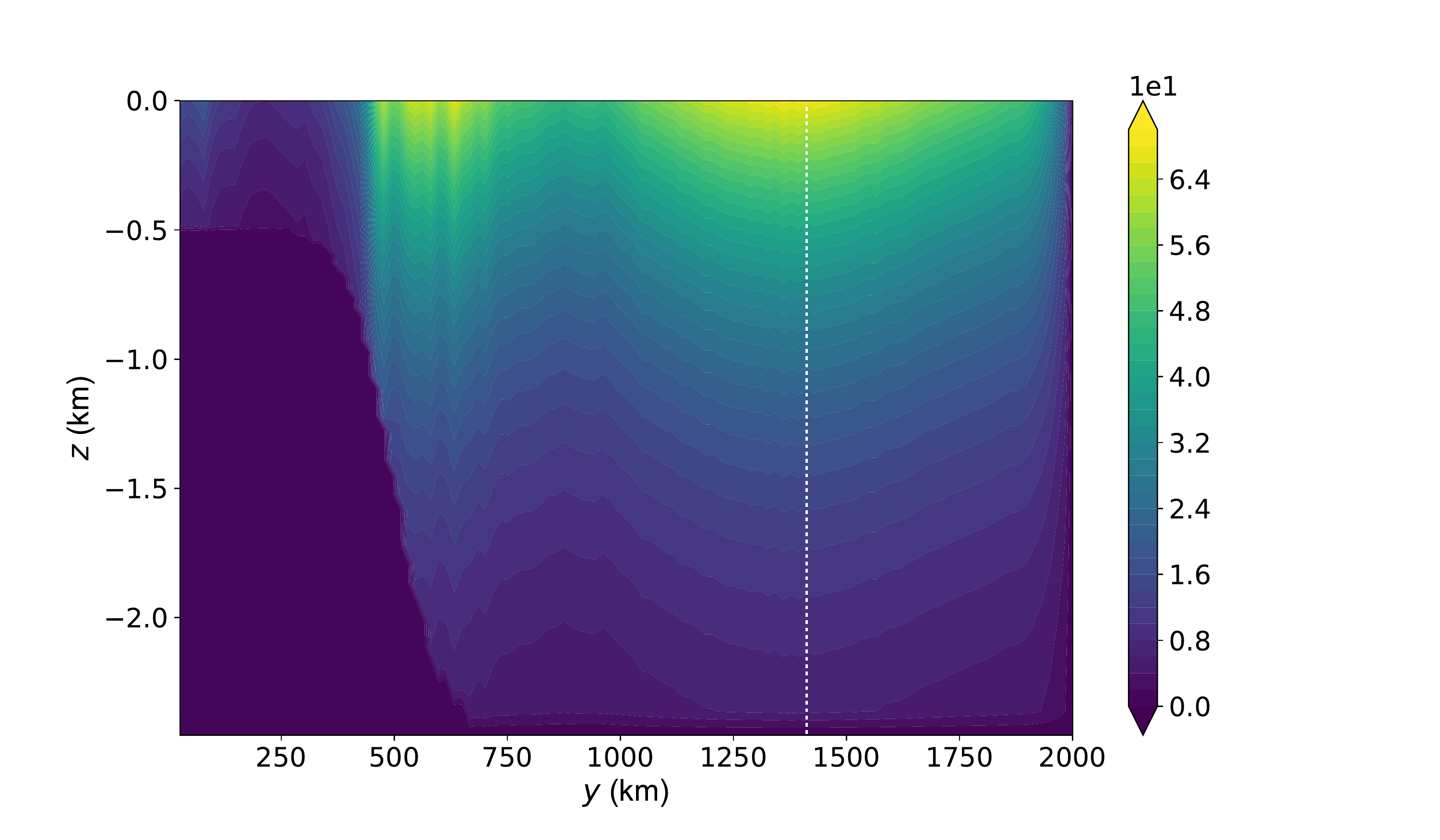}
\caption{Model vertical viscosity (color, in m$^2$/s) from the pZISO simulation.}
\label{fig:visc_pZISO}
\end{figure}

The model vertical viscosity $\mu$ calculated by our algorithm (\ref{eq:visc_in_algorithm}) is plotted in figure \ref{fig:visc_pZISO}.
The viscosity is concentrated within regions where there are large vertical zonal velocity gradients, within the main channel and over the continental shelf, and is largest towards the surface of these two regions, where it peaks at about 65 m$^2$/s.
The viscosity in hrZISO shown in figure \ref{fig:visc_hrZISO}, on the other hand, is largest towards the bottom of the interior region in the main channel, where it peaks at about 20 m$^2$/s.

\begin{figure}[t]
\centering
\noindent\includegraphics[width=0.8\textheight,angle=0]{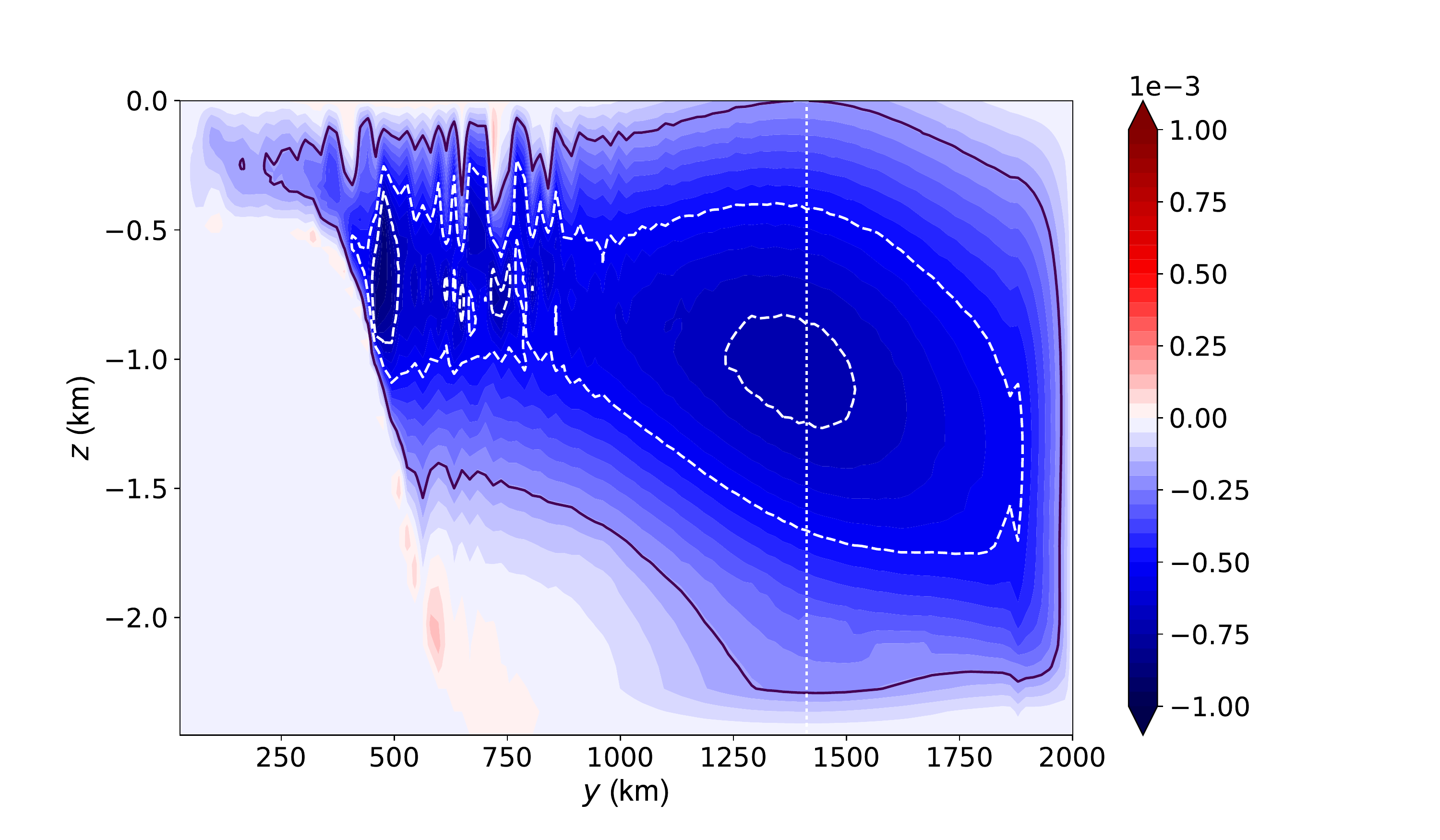}
\caption{Modeled vertical eddy flux of zonal momentum (color, in m$^2$/s$^2$) from the pZISO simulation. The solid black contour line indicates a value of $\sim$ -1.949$\times10^{-4}$ m$^2$/s$^2$, which corresponds to the negative peak wind stress, i.e. -0.2 N/m$^2$; white-dashed lines are contours at -7.0$\times10^{-4}$ and -5.0$\times10^{-4}$ m$^2$/s$^2$.}
\label{fig:vert_flux_xmom}
\end{figure}

The modeled eddy flux of zonal momentum (figure \ref{fig:vert_flux_xmom}) is from the surface towards the bottom of the channel.
At the surface of the circumpolar current in main channel, this flux balances wind stress (0.2 N/m$^2$, solid black line), and becomes larger than $2.0\times10^{-4}$ close to the surface, as it does in the hrZISO between the surface ventilated and interior regions; at the bottom, the flux is balanced by the bottom drag.

\begin{figure}[t]
\centering
\noindent\includegraphics[width=0.8\textheight,angle=0]{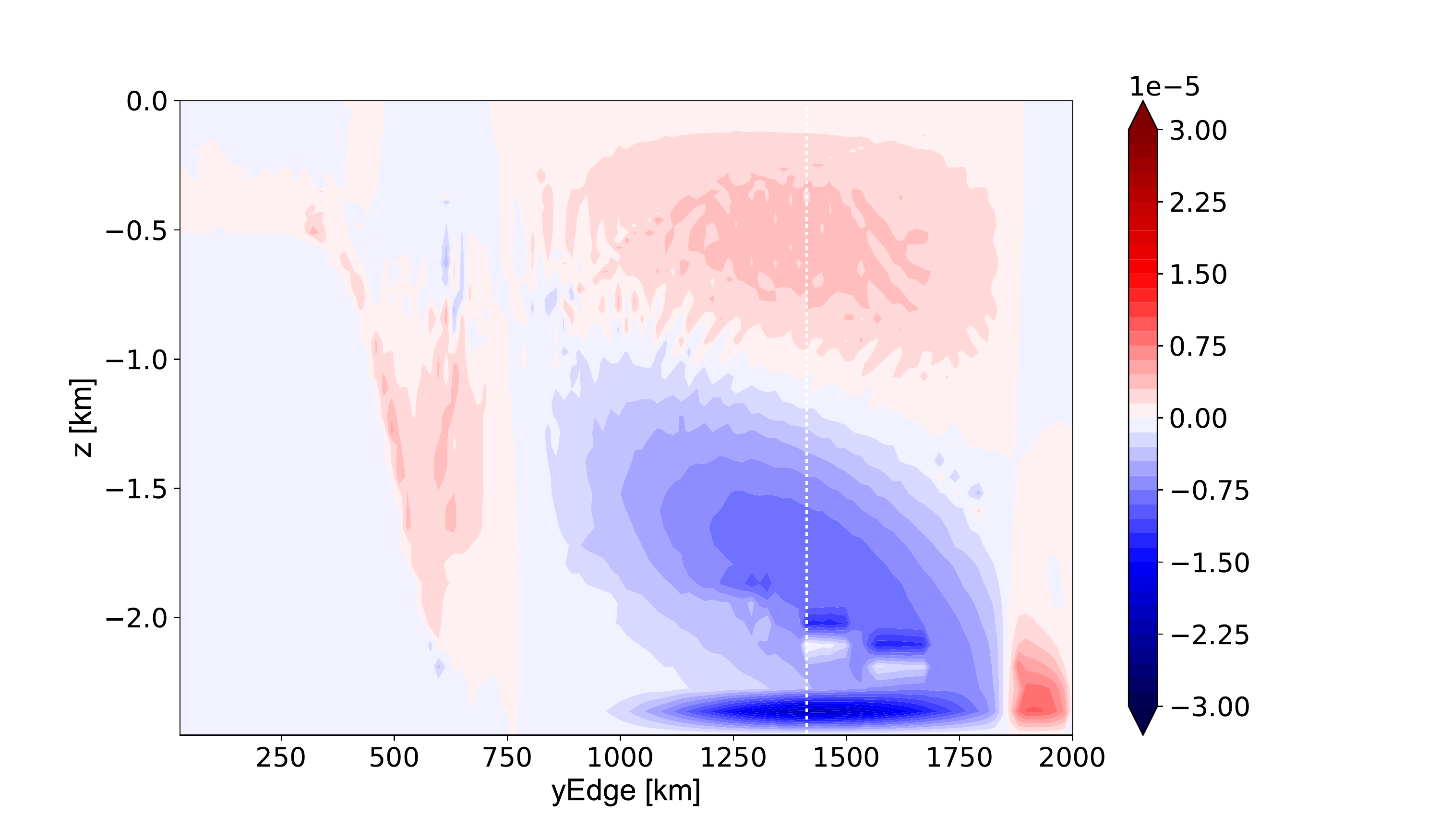}\\
\noindent\includegraphics[width=0.8\textheight,angle=0]{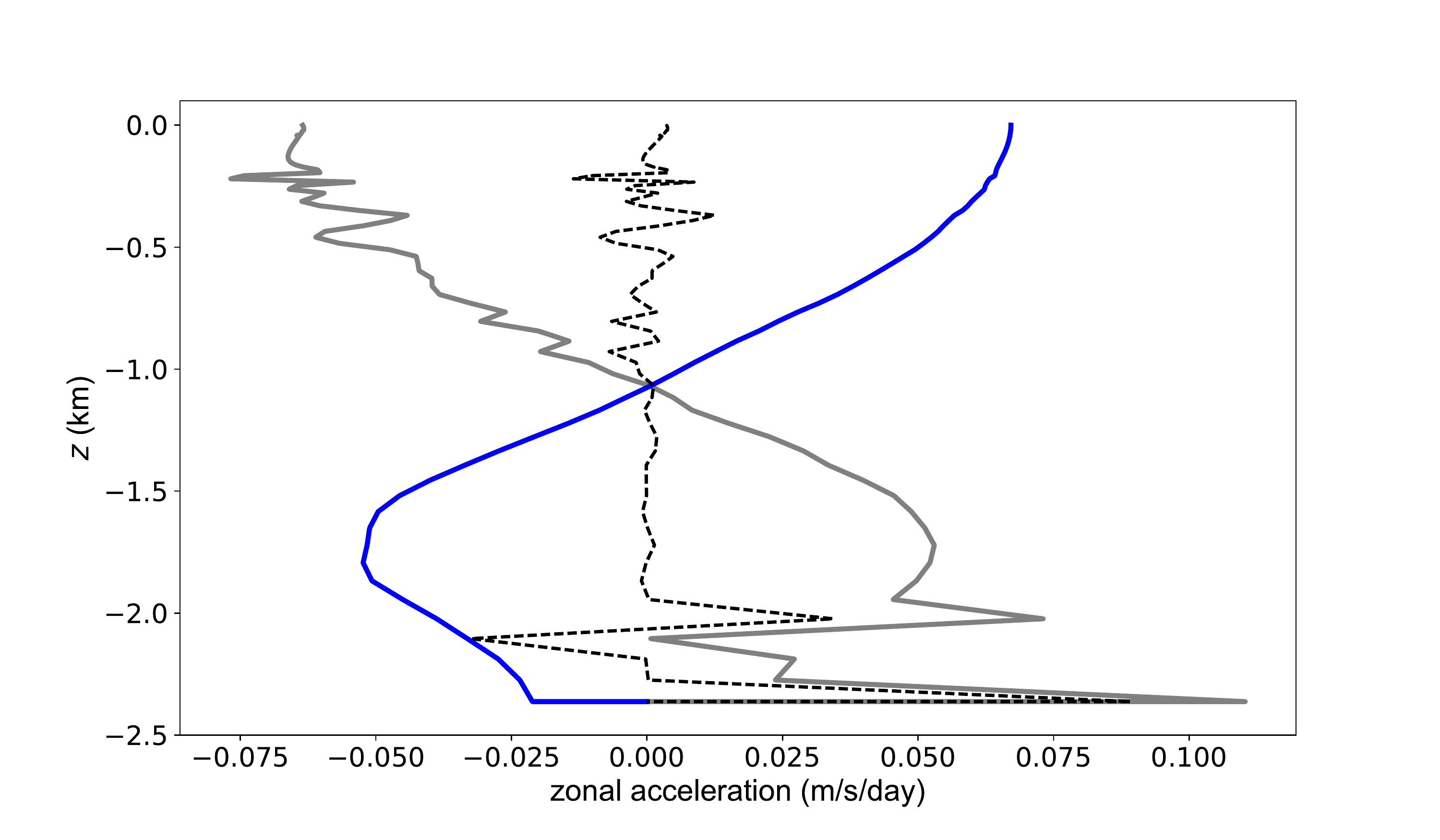}
\caption{Force balance within the eastward circumpolar current of the pZISO in the water column indicated by the vertical white line in figures \ref{fig:par_utwa}-\ref{fig:vert_flux_xmom}: modeled force from eddy form drag (gray), meridional advection of potential vorticity (blue), residual (black-dashed).}
\label{fig:force_balance}
\end{figure}

The force balance within the center of the eastward circumpolar current in the main channel, at the location indicated by the white vertical dashed line in figures \ref{fig:par_utwa}-\ref{fig:vert_flux_xmom}, is plotted in figure \ref{fig:force_balance}.
As in the interior region of the hrZISO simulation, the force balance is between the modeled force from eddy form drag and the advection of potential vorticity by the residual mean meridional velocity $\hat v$.
However, there are important differences between the force balance resulting from the parameterized flow and the one resulting from the eddy resolving flow. 
In upper portion of the interior region in both flows, the eddy forces are negative while the momentum flux from meridional advection is positive.
These forces become zero at a depth of 1 km and 0.6 km in the parameterized and eddy resolving flows, respectively.
Below these depths, the signs of the forces are reversed in the parameterized flow, i.e. the eddy forces become positive while the potential vorticity flux of momentum becomes negative, while in the eddy resolving flow, both become zero.
The main reason for this difference is that, in the parameterized flow, the vertical gradient of the zonal velocity is not negligible, and the modeled vertical eddy flux of zonal momentum is not constant, while in the eddy resolving flow the vertical gradient of the vertical velocity is small and the momentum flux becomes constant.

\section{Discussion and Conclusions}
\label{sec:discussion_conclusions}

We have proposed a model for meso-scale eddy momentum fluxes in the ocean, based on mixing length theory and the gradient diffusion hypothesis.
We diagnosed the model equation in an eddy resolving numerical simulation of a zonally re-entrant channel representative of the Southern Ocean, ZISO.
We have implemented the parmeterization in an ocean model and tested it to simulate a parameterized flow in zonally symmetric idealized Southern ocean configuration.
Our model has several desirable attributes, as well as shortcomings that we shall discuss below.
%Add a segway sentence or two describing the efforts to build in vertical structure into existing parameterizations????
The vertical structure of the model is built in using physical principles, albeit with limitations perhaps due to simplifications and assumptions we have invoked.
However, we have demonstrated the feasibility of using the mathematically consistent, physically-sound thickness-weighted averaged framework and the Eliassen-Palm stress tensor to construct parameterizations with structure and complexities that go beyond the algebraic nature of some existing approaches.

We assume that eddy form drag is the dominating term in the divergence of the Eliassen-Palm stress tensor, and therefor our model is most appropriate for the very simplified case of the zonally symmetric idealized Southern Ocean (ZISO), where this assumption is valid and accurate.
To parameterize the eddy form drag forces from the divergence of the momentum fluxes in the EPFT, we invoke the down gradient diffusion hypothesis to relate these fluxes to the vertical gradient of the zonal TWA velocity via a vertical viscosity.
We relate the vertical viscosity to an eddy velocity proportional to eddy energy, and to an eddy length scale.
The eddy length scale is assumed to be the first Rossby radius of deformation, and the 
Finally, the eddy energy is obtained by using the first baroclinic mode to project the vertically averaged eddy energy onto the water column at a given horizontal location, where the vertically integrated eddy energy is a prognostic variable that is integrated in time with the mean flow.

With our model, we are able to represent some important features and some aspects of the vertical structure of the zonally symmetric flows.
We are able to simulate the vertical flux of zonal momentum from the surface, where it is input by wind stress, to the bottom of the ocean, where it is transferred to bottom drag, as observed in high resolution, eddy resolving simulations \citep{ringler_etal_2017}.
In the interior of the main eastward current, the force balance, given by the divergence of the vertical flux of horizontal momentum, is between eddy forces and the meridional flux of potential vorticity, as in the high resolution simulations.
The fluxes peak at mid depth, consistent with the high resolution flow, but then decrease towards the bottom, while in the high resolution simulations the fluxes are constant and the forces become zero in the bottom half of the water column.
One possible explanation is that in the parameterized flow, the vertical viscosity peaks at the top and decreases with depth, while in the high resolution flow the vertical viscosity in the bottom half of the water column, where the momentum flux is constant.
Thus, we may be over-constraining the vertical structure of the flow with our model for the eigenvector corresponding to the first baroclinic mode, which is used to project the vertically integrated eddy energy onto the water column, and thus gives the vertical viscosity its vertical structure.
The first baroclinic mode is often not the most unstable mode in the ocean \citep{smith_2007}, or the most important in dictating the vertical structure of mixing \citep{smith_marshall_2009}.
We therefore need to investigate the effects of using different eigenvectors, or, alternatively, use a formulation in which the prognostic eddy energy is a function of horizontal location, as well as depth.

%On using the vertically integrated eddy energy, instead of eddy energy. What are we missing by using the vertically integrated ee instead of the full ee.
%On using the first baroclinic mode only, and its model. What are we missing by using it.

The down gradient diffusion has been known to fail in flows where there is no stream-wise symmetry, and where history effects are important.
This is not a problem in the zonally symmetric flow we investigate here, but lack of symmetry and flow history become leading order in the presence of topography, for the flow over or around features such as the Kuerguellan plateu and Drake passage in the Southern ocean, among others.
The down gradient diffusion hypothesis can be bypassed by using a prognostic equation for one or more of the EPFT terms.
For example, for the ZISO case, one could use a prognostic equation for $E_{31}$ and $E_{32}$, and for flows around topography, the diagonal terms $E_{11}$ and $E_{22}$, as we expect that transfer terms between the different components of $E_{ij}$ will become important.
For example, as the flow goes over topography, $E_{11}$ and $E_{22}$ become non-negligible, and as the flow moves away from topography, they may become small again as they transfer momentum to eddy form drag terms $E_{11}$ and $E_{22}$ and/or to the mean flow.

The surface ventilation layer has a leading order effect on the meridional transfer of heat, as well as on the ventilation of the interior of the ocean.
However, we do not include a model for the surface layer in this paper.

\section*{Acknowledgements}
Multiscale. IC. MPAS-O development team.



\appendix

\section{Derivation of the eddy energy equation}
\label{a:eddyenergy}

Using the definitions for mean kinetic energy, mean potential energy, eddy kinetic energy and eddy potential energy in (\ref{eq:def_mke}-\ref{eq:def_epe}), the evolution equations for these quantities can be derived, resulting in
\begin{equation}
\Dts{\Km} 
+ \mathbf{\hat u_h} \cdot \nabla_h \, \ol m 
+  \hat u \nabla \cdot \mathbf{E}^u 
+ \hat v \nabla \cdot \mathbf{E}^v 
= \hat u \hat R_x + \hat v \hat R_y
\label{eq:mke}
\end{equation}
\begin{equation}
\Dts{\Pm} 
- \mathbf{\hat u_h} \cdot \nabla_h \, \ol m 
+ \nabla \cdot \ol m \mathbf{u^\sharp} 
- \nabla \cdot \Pm \mathbf{\hat u_h} 
= \Pm \hat\varpi_{\tilde b}
\label{eq:mpe}
\end{equation}
\begin{eqnarray}
\Dts{\Ke} 
- \hat u \nabla \cdot \mathbf{E}^u - \hat v \nabla \cdot \mathbf{E}^v 
+ \hat u \widehat{m'_{\tilde x} } 
+ \hat v \widehat{m'_{\tilde y} } 
+ \widehat{u'' m'_{\tilde x}}
+ \widehat{v'' m'_{\tilde y}}
\nonumber  \\
+  \nabla \cdot ( \hat u \; \mathbf{J^u} + \hat v \; \mathbf{J^v} ) + \nabla \cdot (\mathbf{J_3^u} + \mathbf{J_3^v}) 
= \widehat{u'' R_x''} + \widehat{v'' R_y''}
\label{eq:eke}
\end{eqnarray}
\begin{eqnarray}
\Dts{\Pe} 
- \hat u \widehat{m'_{\tilde x} } 
- \hat v \widehat{m'_{\tilde y} } 
- \widehat{u'' m'_{\tilde x}}
- \widehat{v'' m'_{\tilde y}}
\nonumber \\
+ \nabla \cdot \ol{ \mathbf{u''_h} m' }
- \nabla \cdot \left \{ \ol{ m'  \mathbf{\hat u _h} \cdot \nabla_h \zeta'  } \; \mathbf{\ol e_3} \right \}
- \nabla \cdot \left \{ \ol{ m'  \mathbf{ u'' _h} \cdot \nabla_h \zeta'  } \; \mathbf{\ol e_3} \right \}
\nonumber \\
- \nabla \cdot \Pe \mathbf{\hat u_h} 
+ \nabla_h \cdot \frac{\mathbf{\hat u_h} \ol{m' \sigma' } }{\ol \sigma} 
+ \nabla_h \cdot \frac{ \ol{ \mathbf{u''_h} m' \sigma' } }{\ol \sigma} 
\nonumber \\
= \Pe \hat\varpi_{\tilde b}
+ \frac{1}{\ol \sigma} \left [ 
   \ol{ m'(\hat \varpi \sigma')_{\tilde b} }
+ \ol{ m'(\varpi'' \sigma')_{\tilde b} }
-  (\hat \varpi \ol{ m' \sigma' } )_{\tilde b} 
\right.
\nonumber \\
\left .
-  (\ol{ \varpi'' m' \sigma' } )_{\tilde b}
- \ol \sigma \ol{ \varpi'' \zeta'}
- \ol{ \sigma' \varpi'' \zeta'} + \hat \varpi \ol{\sigma' \zeta'}
 \right ],
\label{eq:epe}
\end{eqnarray}
where $\nabla_h$ is the horizontal gradient operator, 
$\mathbf{\hat u_h} = \hat u \mathbf{i} + \hat v \mathbf{j}$ the horizontal residual-mean velocity, $\mathbf{u''_h} = u'' \mathbf{i} + v'' \mathbf{j}$ is the horizontal perturbation velocity vector.

Note that in the above equations, $\hat u \widehat{m'_{\tilde x} } $ and $\hat v \widehat{m'_{\tilde y} } $ represent work by the mean flow on eddy pressure gradient, proportional to form drag.

The equation for eddy energy $\Ee = \Ke + \Pe$ becomes
\begin{eqnarray}
\Dts{\Ee} 
- \hat u \nabla \cdot \mathbf{E}^u - \hat v \nabla \cdot \mathbf{E}^v 
\nonumber  \\
+  \nabla \cdot ( \hat u \; \mathbf{J^u} + \hat v \; \mathbf{J^v} ) + \nabla \cdot (\mathbf{J_3^u} + \mathbf{J_3^v}) 
\nonumber  \\
+ \nabla \cdot \ol{ \mathbf{u''_h} m' }
- \nabla \cdot \left \{ \ol{ m'  \mathbf{\hat u _h} \cdot \nabla_h \zeta'  } \; \mathbf{\ol e_3} \right \}
- \nabla \cdot \left \{ \ol{ m'  \mathbf{ u'' _h} \cdot \nabla_h \zeta'  } \; \mathbf{\ol e_3} \right \}
\nonumber \\
- \nabla \cdot \Pe \mathbf{\hat u_h} 
+ \nabla_h \cdot \frac{\mathbf{\hat u_h} \ol{m' \sigma' } }{\ol \sigma} 
+ \nabla_h \cdot \frac{ \ol{ \mathbf{u''_h} m' \sigma' } }{\ol \sigma} 
\nonumber \\
= \widehat{u'' R_x''} + \widehat{v'' R_y''} + \Pe \hat\varpi_{\tilde b}
+ \frac{1}{\ol \sigma} \left [ 
   \ol{ m'(\hat \varpi \sigma')_{\tilde b} }
+ \ol{ m'(\varpi'' \sigma')_{\tilde b} }
-  (\hat \varpi \ol{ m' \sigma' } )_{\tilde b} 
\right.
\nonumber \\
\left .
-  (\ol{ \varpi'' m' \sigma' } )_{\tilde b}
- \ol \sigma \ol{ \varpi'' \zeta'}
- \ol{ \sigma' \varpi'' \zeta'} + \hat \varpi \ol{\sigma' \zeta'}
 \right ],
\label{eq:epe}
\end{eqnarray}
Written more succinctly, 
\begin{equation}
\ddtb{\Ee} 
+ \mathbf{u^\sharp} \cdot \nabla \Ee 
- \hat u \nabla \cdot \mathbf{E}^u 
- \hat v \nabla \cdot \mathbf{E}^v  
+ \nabla \cdot \mathbf{T} 
= \mathcal{D}_e,
\label{eq:eddy_energy}
\end{equation}
where $\mathbf{u^\sharp} = \hat u \mathbf{i} + \hat v \mathbf{j} + w^\sharp \mathbf{k}$ the terms on the left hand side correspond to the time rate of change of eddy energy, transport of eddy energy by the mean flow, eddy-mean flow interactions through the EPFT and eddy energy transport by eddy fluxes, and on the right hand side we have eddy energy dissipation by adiabatic processes.


\section*{References}

\bibliographystyle{elsarticle-harv}
\bibliography{mep_epft}
%bibexport -o mep_epft.bib mep_epft.aux
%\bibliography{master_bibliography}

%%%%%%%%%%%%%%%%%
% Figures
%%%%%%%%%%%%%%%%%

%\listoffigures

\end{document}